\documentclass{elsart5p}
\usepackage[english]{babel}
\usepackage[latin1]{inputenc}
\usepackage[T1]{fontenc}
\usepackage{amssymb}
\usepackage{mathrsfs}
\usepackage[dvips]{graphicx}
\usepackage{epsfig,bm}

\newcommand{\beq}{\begin{equation}}
\newcommand{\eeq}{\end{equation}}

\begin{document}
\begin{frontmatter}

\title{Stabilizing the intensity for a Hamiltonian model of the FEL}

\date{\today}

\author[cpt]{R.~Bachelard}
\author[cpt]{C.~Chandre}
\author[man]{D.~Fanelli}
\author[cpt]{X.~Leoncini}
\author[cpt]{M.~Vittot}

\address[cpt]{Centre de Physique Th\'eorique\thanksref{umr}, CNRS Luminy, Case 907, F-13288 Marseille Cedex 9, France}
\address[man]{Theoretical Physics Group, School of Physics and Astronomy, The University of Manchester, Manchester, M13 9PL, UK}

\thanks[umr]{ Unit\'e Mixte de Recherche (UMR 6207) du CNRS, et des universit\'es Aix-Marseille I, Aix-Marseille II et du Sud Toulon-Var. Laboratoire affili\'e \`a la FRUMAM (FR 2291). Laboratoire de Recherche Conventionné du CEA (DSM-06-35).} 

\begin{abstract}
The intensity of an electromagnetic wave interacting self-consistently
with a beam of charged particles, as in a Free Electron Laser, displays large
oscillations due to an aggregate of particles, called the macro-particle. In
this article, we propose a strategy to stabilize the intensity by destabilizing
the macro-particle. This strategy involves the study of the linear stability
of a specific periodic orbit of a mean-field model. As a control parameter - the amplitude of an external wave - is varied, a bifurcation occur in the
system which has drastic effects on the self-consistent
dynamics, and in particular, on the macro-particle. We show how to obtain an
appropriate tuning of the control parameter which is able to strongly decrease the
oscillations of the intensity without reducing its mean-value.
\end{abstract}


\end{frontmatter}

\section{Introduction}\label{intro}

The amplification of a radiation field by a beam of particles and the radiated field, as it occurs in a Free Electron Laser, can
be modelled within the framework of a simplified Hamiltonian
\cite{bonifacio}. The $N+1$ degree of freedom Hamiltonian
displays a kinetic part, associated with the $N$ particles, and a
potential term accounting for the self-consistent interaction between the
particles and the wave. Thus, mutual particles interactions are neglected, while 
an effective coupling is indirectly provided through the wave.

The linear theory predicts \cite{bonifacio} for the amplitude of the radiation field a linear exponential instability, and then a late oscillating
saturation. Inspection of the asymptotic phase-space suggests that a
bunch of particles gets trapped in the resonance and forms a clump
that evolves as a single {\it macro-particle} localized in phase space. The
untrapped particles are almost uniformly distributed between two
oscillating boundaries, and form the so-called {\it chaotic sea}.

Furthermore, the macro-particle rotates around a well defined centre in phase-space and this
peculiar dynamics is shown to be responsible for the
macroscopic oscillations observed for the intensity \cite{tennyson,antoniazzi}.
It can be
therefore hypothesized that a significant reduction in the intensity
fluctuations can be gained by implementing a dedicated control
strategy, aimed at reshaping the macro-particle in space.

The dynamics can be also investigated from a topological point of
view, by looking at the phase space structures. In the framework of a simplified
mean field description, i.e. the so-called {\it test-particle} picture where the particles passively
interact with a given electromagnetic wave: The trajectories of trapped
particles correspond to invariant tori, whereas
unbounded particles evolve in a chaotic region of phase-space. 
Then, the macro-particle corresponds to a dense set of invariant tori.

For example, a static electric field \cite{tsunoda,morales} can be
used to increase the average wave power. While the chaotic particles
are simply accelerated by the external field, the trapped ones are
responsible for the amplification of the radiation field. Some shift in the relative phase between the electrons and the ponderomotive potential can also be implemented to improve harmonic generation.

In this paper, we propose to perturb the system with external electromagnetic waves. Our strategy is to stabilize the intensity of the
wave, by chaotizing the part of phase-space occupied by the macro-particle. To modify the topology of phase space, an additional test wave is introduced, whose amplitude plays the role of a control parameter. The
residue method
\cite{greene,cary1,artres} is implemented to identify the important local
bifurcations happening in the system when the parameter is varied, by an analysis of
linear stability of a specific periodic orbit. Though first developed in a mean-field approach, our strategy proves to be robust as the self-consistency of the wave is restored.

\section{Dynamics of a single particle}\label{meanfield}

The dynamics of the wave particle interaction, as 
encountered in the FEL, can be described by the following $N$-
body Hamiltonian~\cite{bonifacio}: 
\beq\label{HN}
H_{N}(\{\theta_j,p_j\},\phi,I) = \sum_{j = 1}^{N} \frac{p_j^{2}}{2} - 2 \sqrt{\frac{I}{N}} \sum_{j = 1}^{N} \cos{(\phi+\theta_j)}.
\eeq
It is composed of a kinetic contribution and an interaction term between the particles and the radiation field~: the $(\theta_j,p_j)$ are the conjugate phase and momentum of the $N$ particles, whereas $(\phi,I)$ stand respectively for the conjugate phase and
intensity of the radiation field. Furthermore, there are two conserved quantities~:
$H_N$ and the total momentum $P_N=I+\sum_j p_j$. We consider the
dynamics given by Hamiltonian (\ref{HN}) on a $2N$-dimensional
manifold (defined by $H_N=0$ and $P_N=\varepsilon$ where $\varepsilon$
is infinitesimally small).

\begin{figure} 
  \centerline{
    \mbox{\includegraphics[width=1.8in, height=1.3in]{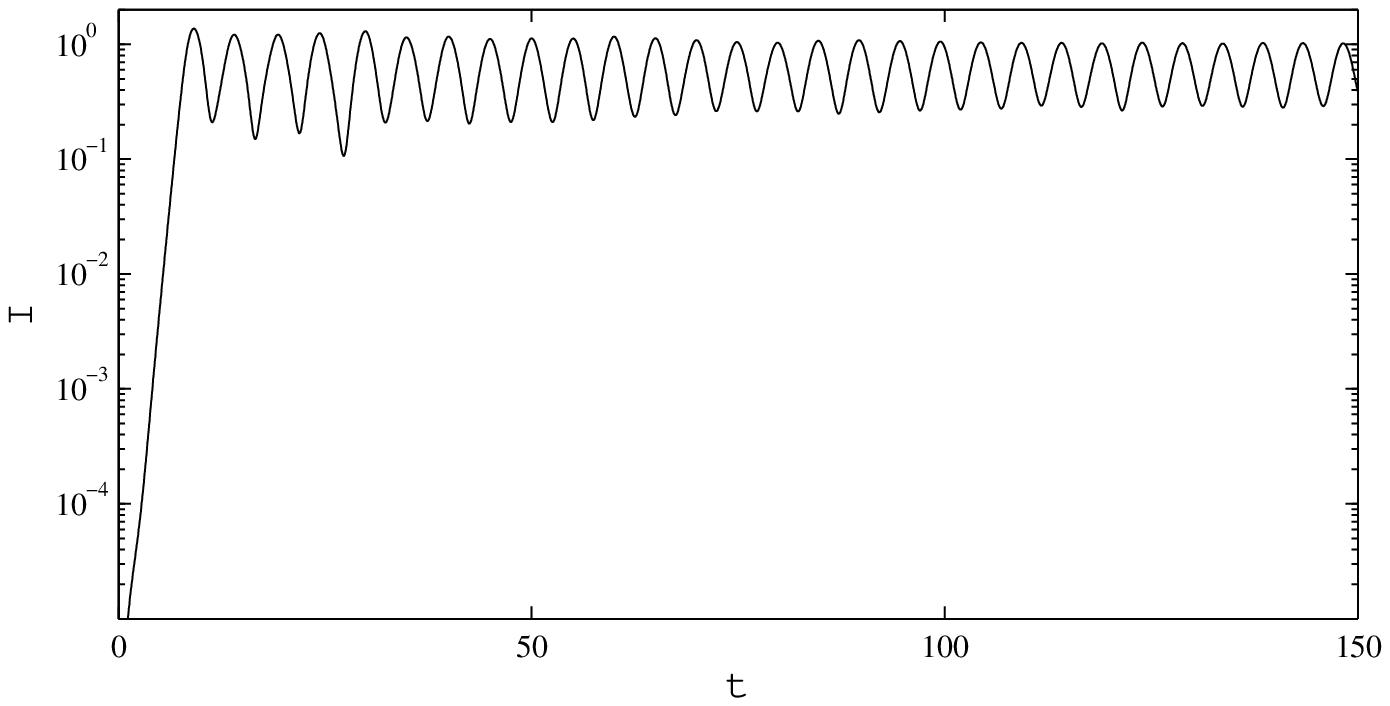}
    \includegraphics[width=1.8in, height=1.3in]{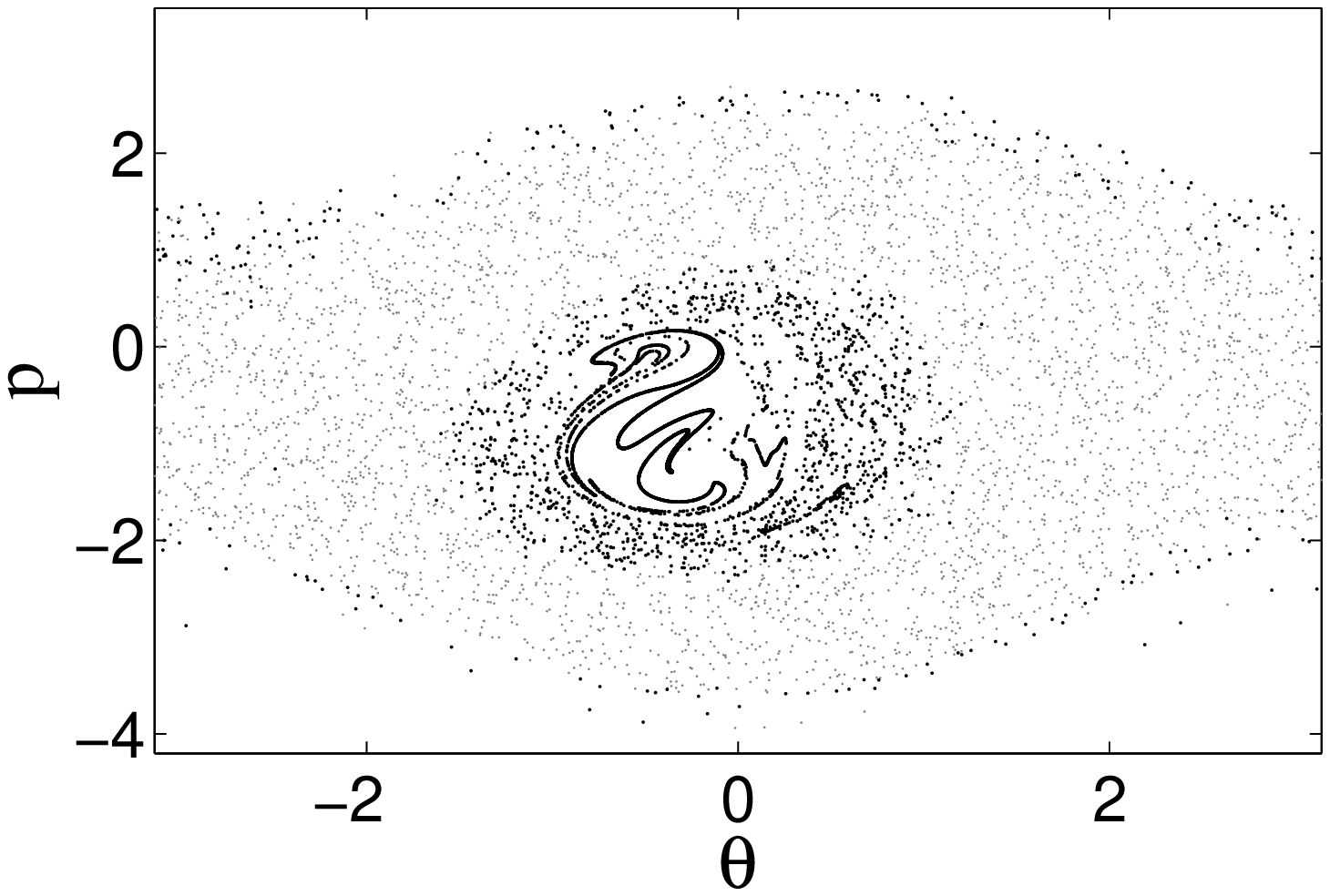}}}
  \caption{Left~: Normalized intensity $I/N$ from the dynamics of Hamiltonian (\ref{HN}), with $N=10000$ particles and $H_N=0$, $P_N=10^{-7}$. Right~: Snapshot of the $N$ particles at $t=800$, with $N=10000$. The grey points correspond to the chaotic particles, the dark ones to the particles in the macro-particle.\label{int0}}
\end{figure}

Starting from a negligible level ($I \ll N$ and $p_j=0$), the intensity grows exponentially and eventually
reaches a saturated state characterized by large oscillations, as depicted in
Fig. \ref{int0}. Concerning the particles dynamics, more than half of them are trapped by the wave \cite{stabilizing} and form the so-called macro-particle (see Fig. \ref{int0}). The remaining particles experience an erratic motion within an oscillating water bag, termed {\it chaotic sea}, which is unbounded in $\theta$ contrary to the macro-particle.

In order to know how many particles have a regular motion, we compute finite Lyapunov exponents for each trajectory (the particles are then considered as evolving in an external field). The Lyapunov exponents were computed over a time period of $T=300$ (once the stationary state reached), and a trajectory is considered to be regular if the Lyapunov exponent is smaller than $0.025$ (while it is typically of order $1$ in the chaotic sea).

In order to get a deeper insight into the dynamics, we consider the motion of a
single particle. For large $N$, we assume that its influence on the wave is negligible, thus it can be described as a passive 
particle in an oscillating field. The
motion of this test-particle is described by the  one and a half degree of freedom Hamiltonian~:

\begin{eqnarray}\label{H1P}
H_{1p} (\theta,p,t) & = & \frac{p^{2}}{2} - 2 \sqrt{\frac{I(t)}{N}} \cos{( \theta +\phi(t))}\nonumber \\
& = & \frac{p^{2}}{2} - Re(h(t) e^{i \theta}),
\end{eqnarray}
where the interaction term $h(t)$ is derived from dedicated simulations of the
original self-consistent $N$-body Hamiltonian (\ref{HN}). In the saturated regime, $h(t)$ is mainly periodic. In particular, a refined Fourier analysis shows that it can be written as~:
\begin{equation}\label{inter}
h(t) = 2 \sqrt{\frac{I(t)}{N}} e^{i \phi(t)} \approx [F + \alpha e^{i \omega_1 t} + \beta e^{-i \omega_1 t}] e^{i \Omega t},
\end{equation}
where $\Omega=-0.685$ stands for the wave velocity and $\omega_1=1.291$ for the
frequency of the oscillations of the intensity. As for the amplitudes, the Fourier analysis provides the following values~: $F=1.5382 - 0.0156i$, $\alpha=0.2696 - 0.0734i$ and $\beta=0.1206 + 0.0306i$.

Hamiltonian~(\ref{H1P}) results from a periodic perturbation of a pendulum described by the integrable Hamiltonian $H_0$
$$
H_0=\frac{p^2}{2}-\vert F\vert \cos(\theta+\Omega t+\phi_F),
$$
where $F=\vert F\vert e^{i\phi_F}$. The linear frequency of this pendulum is $\sqrt{\vert F\vert}\approx 1.240$ which is very close to the frequency of the forcing $\omega_1$. Therefore a chaotic behaviour is expected when the perturbation is added even with small values of the parameters $\alpha$ and $\beta$.

\begin{figure}[t] 
  \centerline{
    \mbox{\includegraphics[width=1.8in,height=1.3in]{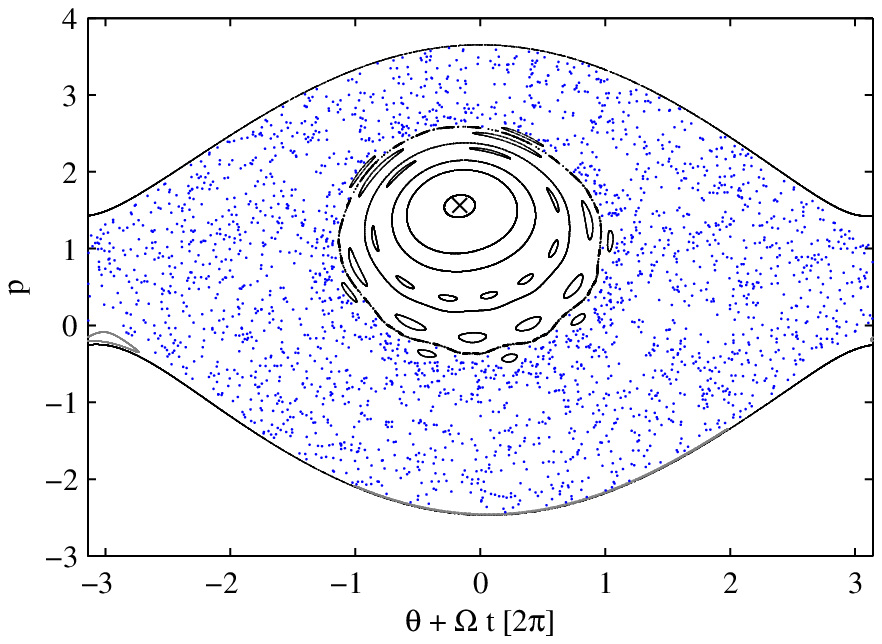}
    \includegraphics[width=1.8in,height=1.3in]{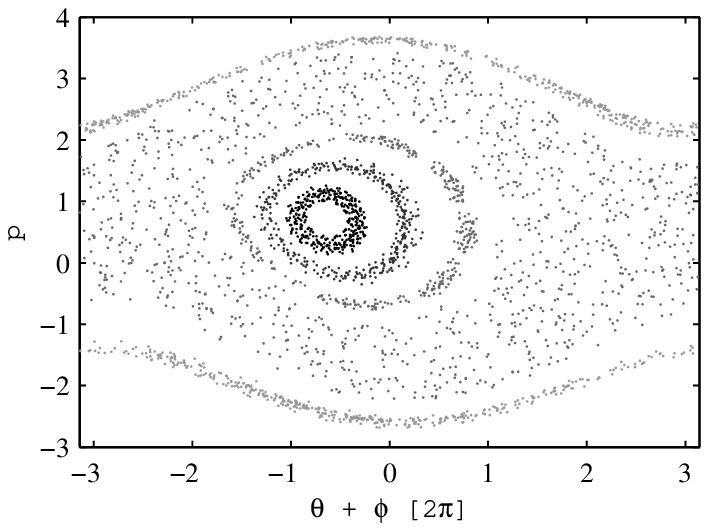}}}
  \caption{Left~: Poincaré section of a test-particle, described by Hamiltonian (\ref{H1P}). The periodic orbit with rotation number $1$ is marked by a cross. Right~: Poincar\'e section of Hamiltonian (\ref{HN}), when the particles intersect the plane $dI(t)/dt=0$. The different trajectories are represented by different grey levels.\label{QPc0}}
\end{figure}

The Poincaré sections (stroboscopic plot performed at frequency
$\omega_1$) of the test-particle (see Fig. \ref{QPc0}) reveal that the
macro-particle reduces to a set of invariant
tori in this mean-field model. Conversely, the chaotic sea is filled with seemingly erratic trajectories
of particles, apart from the upper and lower boundaries, where the trajectories are similar to the rotational ones of the unperturbed pendulum. The rotation of the macro-particle and the
oscillations of the water bag are visualized by translating
continuously in time the stroboscopic plot of phase space. 

The macro-particle is organized around a central (elliptic) periodic orbit with rotation number $1$. The period of oscillations of the intensity is the same as the one of the macro-particle which indicates the role played by this coherent structure in the oscillations of the wave. 

Thus, in the test-particle model, the macro-particle is formed by particles which are trapped on two-dimensional invariant tori. This picture can be extended to the self-consistent model, if one considers the projection of a trajectory $(\phi(t),I(t),\{ \theta_j (t), p_j (t) \}_j)$ in the $(\theta,p)$ plane, each time it crosses the hyperplane $ \sum_{j} \sin{(\phi+\theta_j)} = 0$, i.e. $dI/dt=0$. From the full trajectory, we follow a given particle (an index $j$) and plot $(\theta_j,p_j)$ each time the full trajectory crosses the Poincar\'e section.

The trapped particles appear to be confined to domains of phase-space much smaller than the one of the macro-particle (see Fig.\ref{QPc0}). These domains are similar to the invariant tori of the test-particle model, although thicker.
It is worth noticing that not only these figures have a similar overall layout, but there is a deeper correspondence in the structure of the macro-particle. For instance, both figures show similar resonant islands at the boundary of the regular region. Since we saw that the macro-particle directly influences the oscillations of the wave, the test particle Hamiltonian (\ref{H1P}) serves as a cornerstone of our control strategy which consists in destabilizing the regular structure of the macro-particle in order to stabilize the intensity of the wave. This strategy focuses on breaking up invariant tori to reshape the macro-particle. In order to act on invariant tori, we use the central periodic orbit which, as we have seen, structure the motion of the macro-particle.

\section{Residue method}\label{rescrit}

The topology of phase space can be investigated by analysing the
linear stability of periodic orbits. Information on the nature of these orbits (elliptic,
hyperbolic or parabolic) is provided using, e.g., an indicator like Greene's residue
\cite{greene,mackay}, a quantity that enables to monitor local changes
of stability in a system subject to an external perturbation 
\cite{cary1,artres}.

 From the integration of the equations of the tangent flow of the system along a particular periodic orbit, one can deduce the residue $R$ of this periodic orbit. In particular, if $R\in ]0,1[$, the periodic orbit is called elliptic (and is in general stable); if
$R<0$ or $R>1$ it is hyperbolic; and if $R=0$ and $R=1$, it is parabolic while higher order expansions give the stability of such periodic orbits.

 Since the periodic orbit and its stability depend on the set of parameters $\bm{\lambda}$, the features of the dynamics will change under apposite variations of such parameters. Generically, periodic orbits and their (linear or non-linear) stability properties are robust to small changes of parameters, except at specific values when
 bifurcations occur. The residue method \cite{cary1,artres} detects the rare events where the linear stability of a given periodic orbit changes thus allowing one to calculate the appropriate values of
the parameters leading to the prescribed behaviour of the dynamics. As a consequence, this method can 
 yield reduction as well as enhancement of chaos.

\section{Destruction of the macro-particle}\label{implement}

 The residue method can be used to enlarge the macro-particle in the chaotic sea \cite{stabilizing}, which results in its stabilization~: then, the fluctuations of the intensity of the wave eventually collapse. 
 
 Nonetheless, it can also be used to reduce the aggregation process for the particles, by destroying the invariant tori forming the macro-particle~: such a control, as we will see, tends to limit the fluctuations in the intensity of the wave. Here, we implement this control with an extra test-wave, whose amplitude is used as a control parameter. The Hamiltonian of the mean-field model with a test-wave is chosen as~:
\begin{equation}\label{htestc}
H_{1p}^{c} (\theta,p,t;\lambda) = H_{1p} (\theta,p,t) - 2 \lambda \cos{(k(\theta - \omega_1 t ))},
\end{equation}
where $\omega_1$ corresponds to the resonant frequency of the central periodic orbit of the macro-particle, and $k=10$.
 
\begin{figure}[t!] 
  $\begin{array}{c}
    \mbox{\includegraphics[height=1.3in]{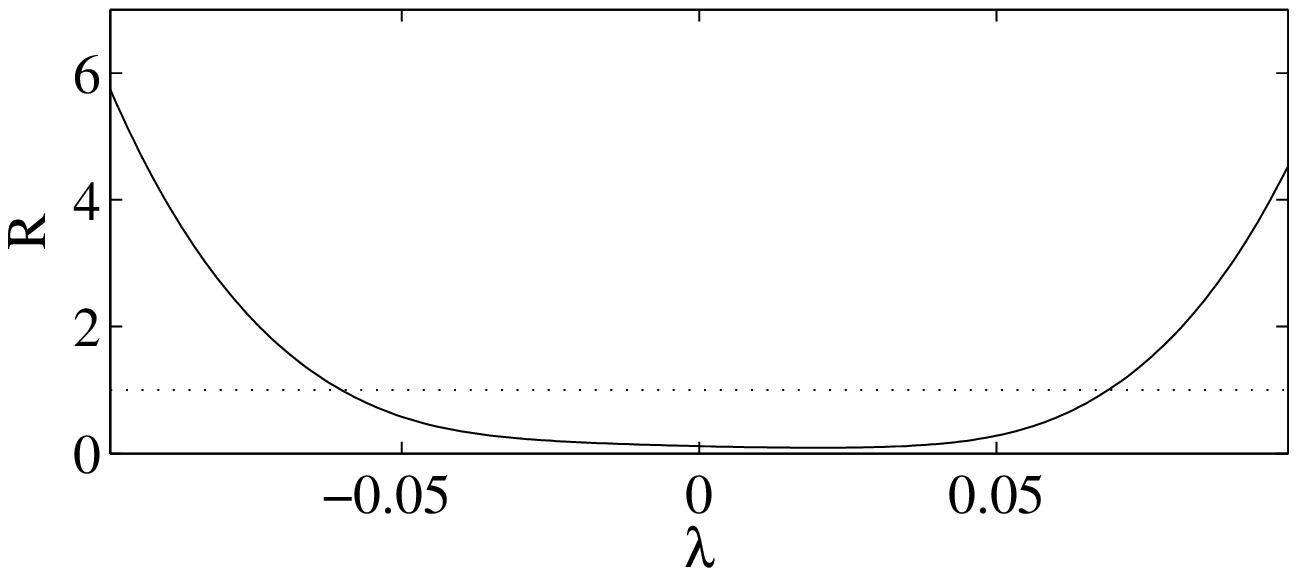}} \\
    \mbox{\includegraphics[width=1.8in,height=1.3in]{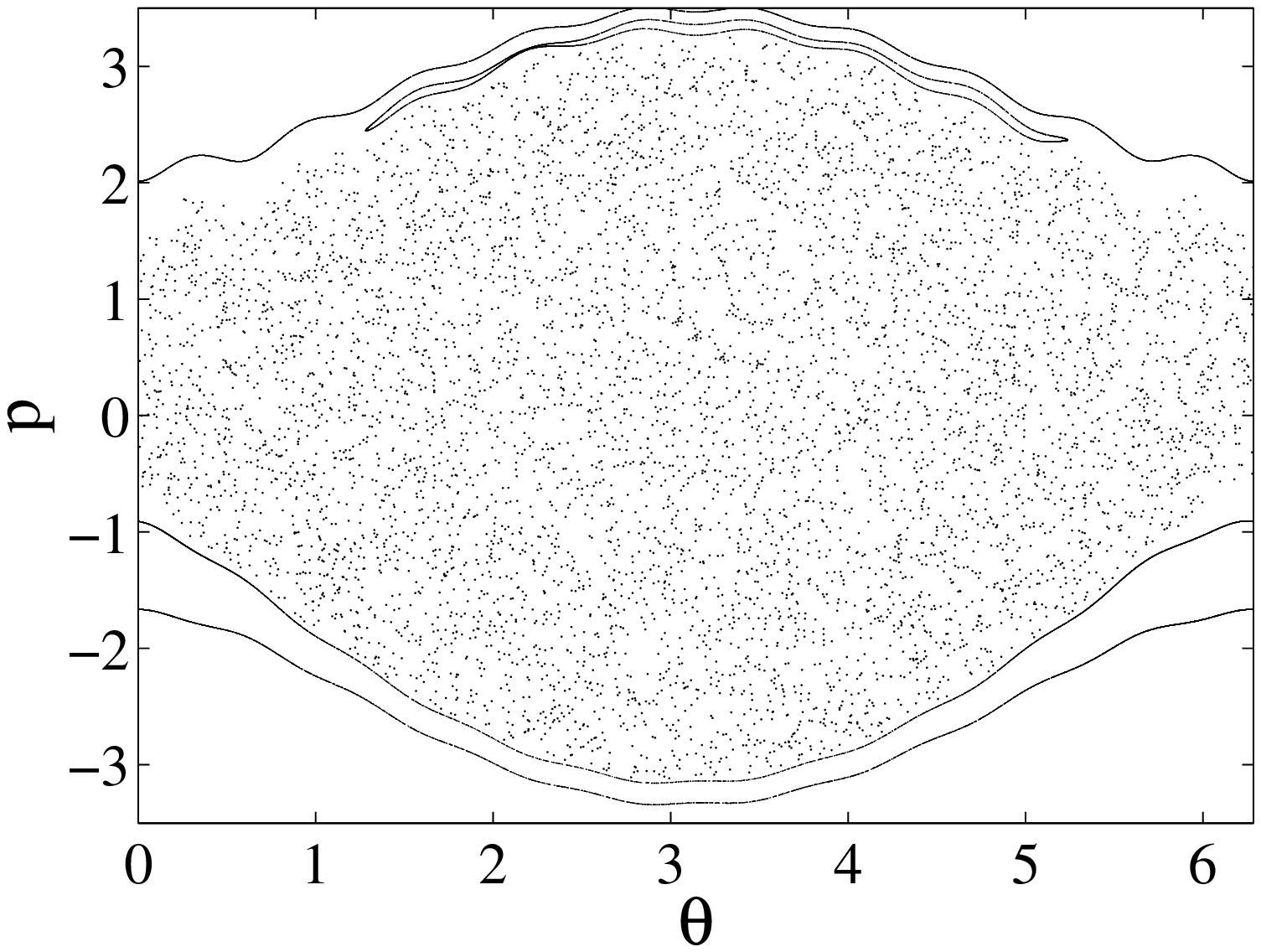}}
   \mbox{\includegraphics[width=1.8in,height=1.3in]{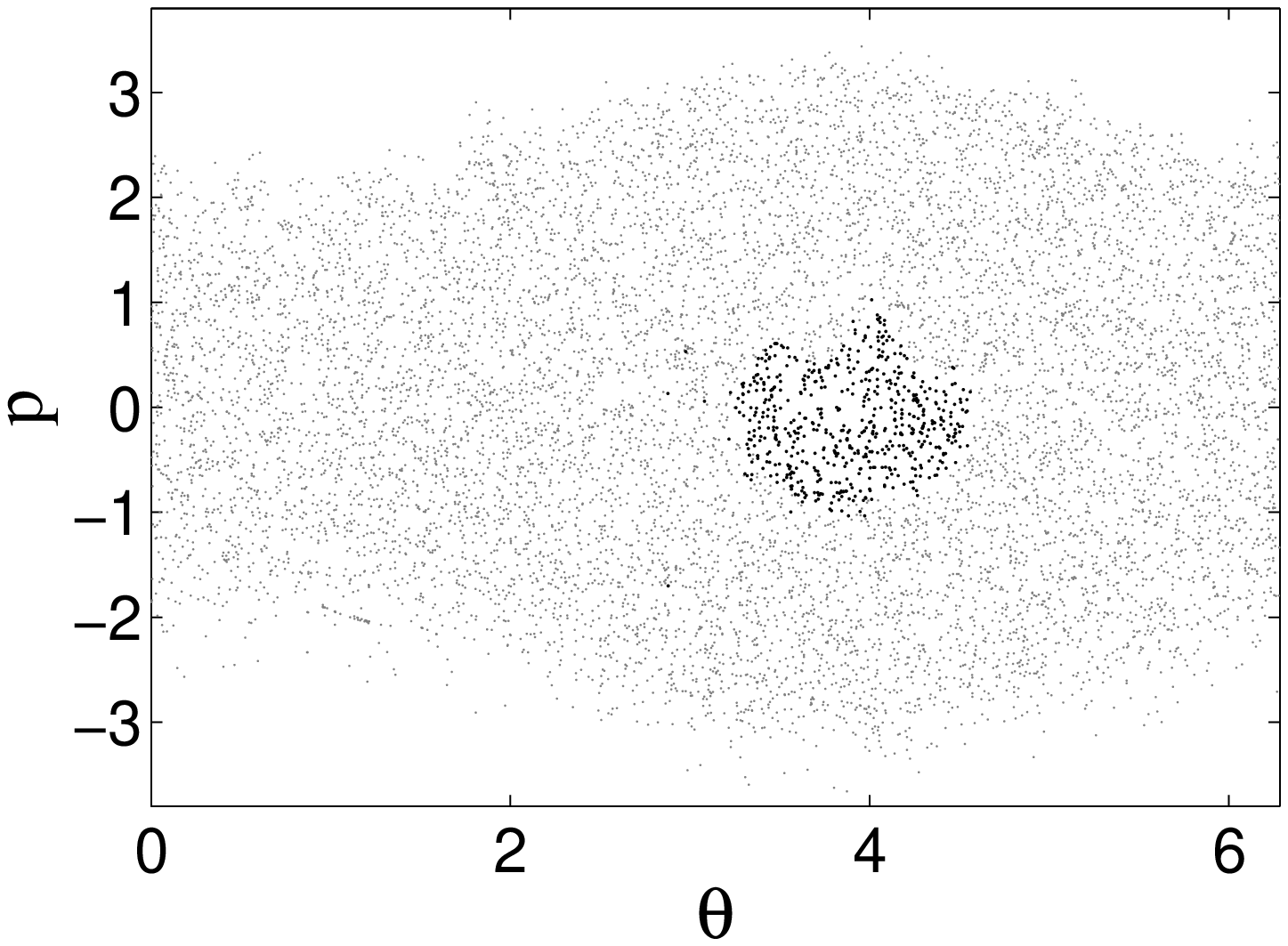}}
  \end{array}$
  \caption{Upper panel~: Residue curve of the periodic orbit of rotation number $1$, as a function of the control parameter $\lambda$. Lower left panel~: Poincaré section of the controlled Hamiltonian (\ref{htestc}) of a test-particle. Lower right panel~: Snapshot of the phase-space of the particles for Hamiltonian (\ref{hn2}), with $N=10000$ and $\lambda=\lambda_c$ (same initial conditions as for Fig.\ref{int0}).\label{figres}}
\end{figure}

Then, the amplitude $\lambda$ is tuned around $0$, and the residue $R$ of the central periodic orbit ${\mathcal O}_1$ is tracked (see Fig.\ref{figres})~: when the latter goes above $1$, it means that the central orbit turned hyperbolic, and that chaos might have locally appeared. This occurs for values of $| \lambda |$ larger than $\lambda_c \approx 0.07$. An inspection of the Poincar\'e section confirms this prediction, as there is no more island with a central periodic orbit of period $2 \pi / \omega_1$. Actually, no more elliptic island can be detected, apart from the borders of the water bag~: thus, though the hyperbolicity of ${\mathcal O}_1$ only guarantees local chaos, the resonance is now fully chaotic, which emphasizes that the study of a few periodic orbits may give quite global information on the dynamics.

 This control strategy can then be generalized to the self-consistent interaction, by introducing a test-wave similar to (\ref{htestc}) in the original $N$-particle Hamiltonian (\ref{HN})~:
\begin{eqnarray}\label{hn2}
H_{N}^{c}(\{\theta_j,p_j\},\phi,I;\lambda) =& H_{N}(\{\theta_j,p_j\},\phi,I) \nonumber \\ & - \lambda_c \sum_j  \cos{( k(\theta_j - \omega_1 t))},
\end{eqnarray}

 Though the control dedicated to the mean-field model lost some of its relevance, due to the presence in the original model of the feedback of the electrons on the wave, the controlled dynamics of the particles is qualitatively similar to the one obtained in the mean-field framework. After an initial growth of the wave, the particles organize themselves in a water bag, but only few of them still display a regular trajectory~: from $65\%$ in the uncontrolled regime, the ratio has collapsed to about $6\%$ for $\lambda=\lambda_c$ (cf Fig.\ref{figres}).  As for the wave, the intensity rapidly stabilizes, after the initial growth. The relevance of a control based on a modification of the macro-particle is thus confirmed. This is in agreement with the experimental results of Dimonte \cite{dimonte}, who observed that one could destroy the oscillations of the intensity with unstable test-waves.

\begin{figure}[t] 
  \begin{center}
    \mbox{\includegraphics[width=1.8in, height=1.2in]{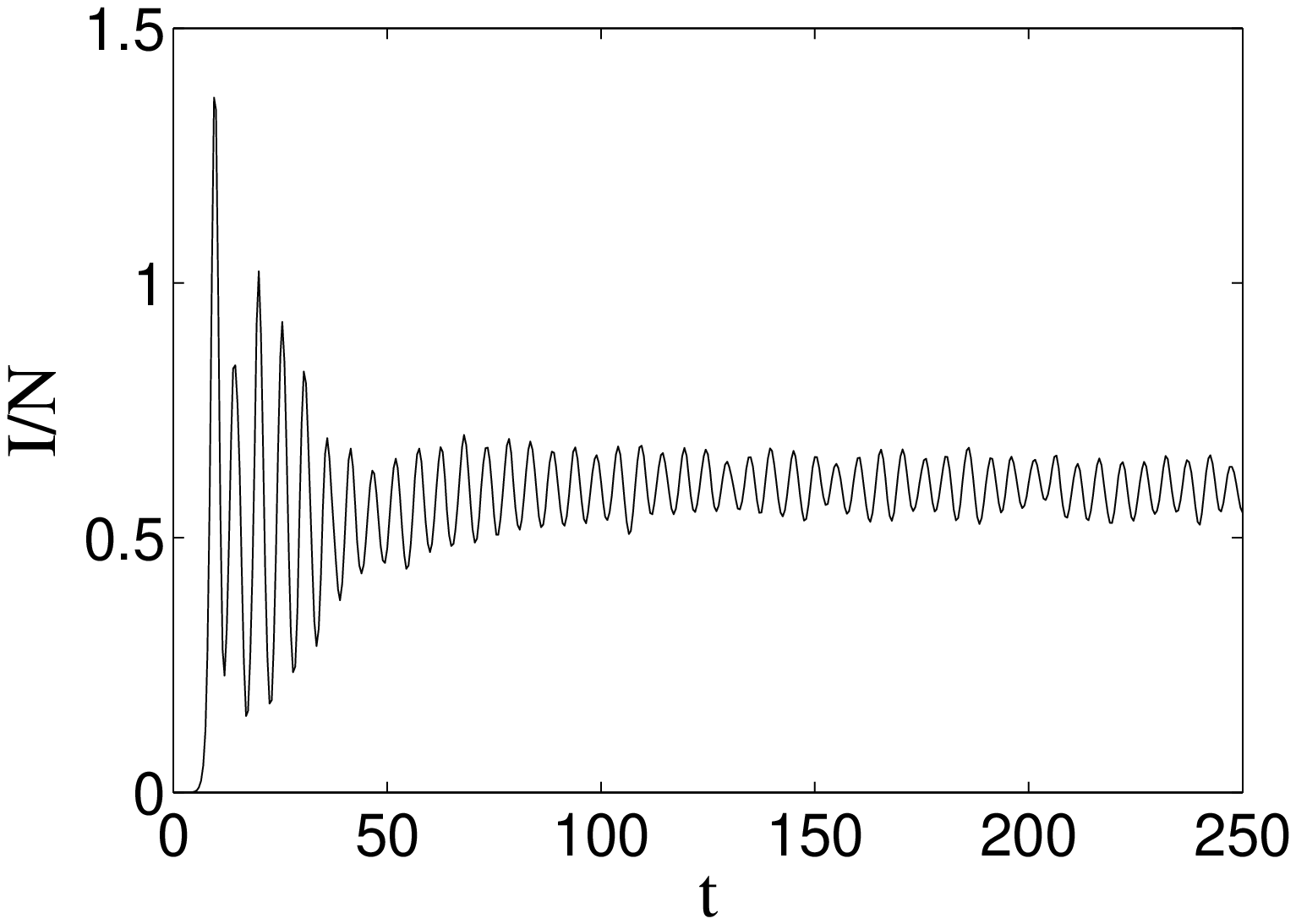}
    \includegraphics[width=1.8in, height=1.2in]{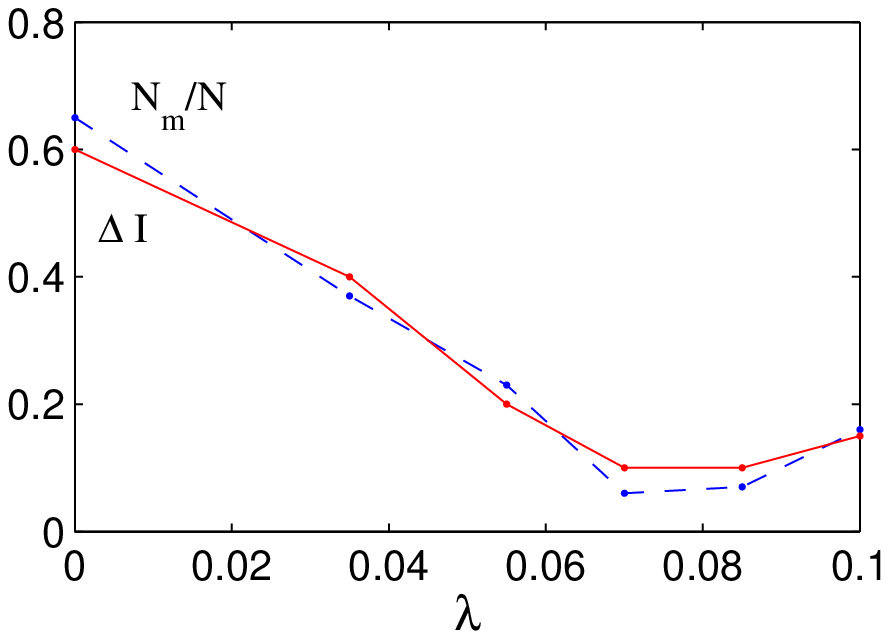}  
    }\end{center}
  \caption{Left~: Normalized intensity $I/N$ for Hamiltonian (\ref{hn2}). Right~: Ratio $N_m/N$ of particles with regular trajectories, for Hamiltonian (\ref{hn2}), as a function of the control parameter $\lambda$. $\Delta I$ corresponds to the mean fluctuations of the intensity.\label{fighnc}}
\end{figure}

Finally, let us note that controlling with a weaker test-wave ($\lambda\leq 0.07$) only partially chaotizes the macro-particle~: the intensity of the wave still stabilizes, though not as much as for $\lambda=\lambda_c$ (see Fig.\ref{fighnc}). Then, a stronger test-wave does not provide a better control, due to the creation of new resonance islands in the test-particle phase space for larger $\lambda$.

\section*{Conclusion}

We proposed in this paper a method to stabilize the intensity of a wave amplified by a beam of particles. This is achieved by destroying the coherent structures of the particles dynamics. By studying a mean-field version of 
the original Hamiltonian setting and putting forward an analysis of 
the linear stability of the periodic orbit, we were able to enhance 
the degree of mixing  of the system: Regular trajectories are turned 
into chaotic ones as the effect of a properly tuned  test-wave, which is 
externally imposed. The results are then translated into the 
relevant $N$-body self consistent framework allowing us to conclude 
upon the robustness of the proposed control strategy.

\section*{Acknowledgements}

This work is supported by Euratom/CEA (contract EUR~344-88-1~FUA~F) and GDR n°2489 DYCOEC. 
We acknowledge useful discussions with G.~De~Ninno, Y.~Elskens and the Nonlinear Dynamics group at Centre de Physique Th\'eorique.


\begin{thebibliography}{}

\bibitem{bonifacio}
R.~Bonifacio, {\it et al.}, Rivista del Nuovo Cimento \textbf{3}, 1 (1990)

\bibitem{tennyson}
J.L.~Tennyson, J.D.~Meiss, and P.J.~Morrison, Physica D \textbf{71}, 1 (1994)

\bibitem{antoniazzi}
A.~Antoniazzi, Y.~Elskens, D.~Fanelli and S.~Ruffo, Europ. Phys. J. B \textbf{50}, 603 (2006)

\bibitem{tsunoda}
S.I.~Tsunoda, J.H.~Malmberg, Phys. Rev. Lett. \textbf{49}, 546 (1982)

\bibitem{morales}
G.J.~Morales, Phys. Fluids \textbf{23} (1980) 

\bibitem{greene}
J.M.~Greene, J. Math. Phys. \textbf{20}, 1183 (1979)

\bibitem{cary1}
J.R.~Cary, J.D.~Hanson, Phys. Fluids \textbf{29}, 2464 (1986)

\bibitem{artres}
R.~Bachelard, C.~Chandre, X.~Leoncini, Chaos \textbf{16}, 023104 (2006)

\bibitem{stabilizing}
R.~Bachelard, A.~Antoniazzi, C.~Chandre, D.~Fanelli, X.~Leoncini, M.~Vittot, Eur. Phys. J. D, \textbf{42}, 125 (2007)

\bibitem{mackay}
R.S.~MacKay, Nonlinearity \textbf{5}, 161 (1992)

\bibitem{dimonte}
G.~Dimonte, and J.H.~Malmberg, Phys. Rev. Lett. \textbf{38}, 401 (1977)

\end{thebibliography}
\end{document}